\documentclass[twocolumn,showpacs,showkeys,amsmath,amssymb,nofootinbib,superscriptaddress,floatfix]{revtex4}

\usepackage{graphicx}
\usepackage{subfigure}
\usepackage{bm}

\def\vec#1{\mathchoice{\mbox{\boldmath$\displaystyle#1$}}
{\mbox{\boldmath$\textstyle#1$}}
{\mbox{\boldmath$\scriptstyle#1$}}
{\mbox{\boldmath$\scriptscriptstyle#1$}}}
\makeatletter
\newcommand\erfc{\mathop{\operator@font erfc}\nolimits}
\def\slashchar#1{\setbox0=\hbox{$#1$}
   \dimen0=\wd0 \setbox1=\hbox{/} \dimen1=\wd1
   \ifdim\dimen0>\dimen1 \rlap{\hbox to \dimen0{\hfil/\hfil}} #1
   \else  \rlap{\hbox to \dimen1{\hfil$#1$\hfil}} / \fi}
\def\p{\slashchar{p}}
\def\w{\omega}

\newcommand{\lrd}{\raisebox{0.09em}{$\stackrel{\scriptstyle\leftharpoonup\hspace{-.7em}\rightharpoonup}{D} $}}
\newcommand{\ld}{\raisebox{0.09em}{$\stackrel{\scriptstyle\leftharpoonup}{D}$}}
\newcommand{\rd}{\raisebox{0.09em}{$\stackrel{\scriptstyle\rightharpoonup}{D}$}}

\newcommand{\ket}[1]{\left| #1 \right>}
\newcommand{\bra}[1]{\left< #1 \right|}

\newcommand{\eVdist}{\kern-0.06667em}

\makeatother

\begin{document}
 
\title{Gravitational and higher-order form factors of the pion in
  chiral quark models\footnote{Supported by Polish Ministry of Science
    and Higher Education, grants N202~034~32/0918 and N~N202~249235, Spanish DGI and
    FEDER funds with grant FIS2005-00810, Junta de Andaluc{\'\i}a
    grant FQM225-05, and EU Integrated Infrastructure Initiative
    Hadron Physics Project contract RII3-CT-2004-506078.}}

\author{Wojciech Broniowski} \email{Wojciech.Broniowski@ifj.edu.pl}
\affiliation{The H. Niewodnicza\'nski Institute of Nuclear Physics,
  Polish Academy of Sciences, PL-31342 Krak\'ow, Poland}
\affiliation{Institute of Physics, Jan Kochanowski University,
  PL-25406~Kielce, Poland} \author{Enrique Ruiz Arriola}
\email{earriola@ugr.es} \affiliation{Departamento de F\'{\i}sica
  At\'omica, Molecular y Nuclear, Universidad de Granada, E-18071
  Granada, Spain}

\date{10 September 2008}

\begin{abstract}
The gravitational form factor of the pion is evaluated in two chiral
quark models and confronted to the recent full-QCD lattice data. We
find good agreement for the case of the Spectral Quark Model, which
builds in the vector-meson dominance for the charge form factor.  We
derive a simple relation between the gravitational and electromagnetic
form factors, holding in the considered quark models in the chiral
limit. The relation implies that the gravitational mean squared radius
is half the electromagnetic one. We also analyze higher-order
quark generalized form factors of the pion, related to higher moments
in the symmetric Bjorken $X$-variable of the generalized parton
distribution functions, and discuss their perturbative QCD evolution,
needed to relate the quark-model predictions to the lattice data. The
values of the higher-order quark form factors at $t=0$, computed on
the lattice, also agree with our quark model results within the
statistical and method uncertainties.
\end{abstract}

\pacs{12.38.Lg, 11.30, 12.38.-t}

\keywords{pion gravitational form factor, generalized parton
  distributions, generalized form factors, structure of the pion,
  chiral quark models}

\maketitle 

\section{Introduction \label{sec:intro}}

Form factors carry basic information on the extended structure of
hadrons as they correspond to matrix elements of conserved currents between hadron states. 
There exist abundant experimental data
concerning the charge form factor, related to the electromagnetic current
and providing the distribution of charge in a hadron.  The {\em
  gravitational form factors}, related to matrix elements of the
energy-momentum tensor~\cite{PhysRev.144.1250} in a hadronic state and
thus providing the distribution of matter within the hadron, are not
experimentally known. This is because whereas the electromagnetic
interactions are probed by structureless electrons and mediated by 
one-photon exchange, the parallel one-graviton exchange is extremely weak
and impossible to measure.  Recently, however, these objects were determined {\it
  ab initio} and with sufficient accuracy in full-QCD lattice simulations
by the QCDSF/UKQCD collaboration \cite{Brommel:2005ee,Brommel:PhD}.
In this paper we confront the predictions of chiral quark models for
the pion gravitational form factors and higher-order generalized form
factors to the lattice determination of
Refs.~\cite{Brommel:2005ee,Brommel:PhD}. We find that in the Spectral
Quark Model (SQM) \cite{RuizArriola:2003bs}, which is a variant of a
chiral quark model with built-in vector meson dominance, the agreement
with the lattice data is especially good. In particular, a slower
fall-off with the large space-like momenta than for the case of the
electromagnetic form factor is found.  We also perform the
calculations in the Nambu--Jona-Lasinio (NJL) model with the
Pauli-Villars (PV) regularization, where the agreement is not as good
as in SQM.  In addition, we find that the higher-order generalized
form factors (see below for a definition) at $t=0$ provided by the
full-QCD lattice simulations are properly reproduced in chiral quark
models.

The form factors are related via sum rules to more general objects,
the generalized parton distributions (GPDs) of the pion (for extensive
reviews see
e.g.~\cite{Ji:1998pc,Radyushkin:2000uy,Goeke:2001tz,Bakulev:2000eb,Diehl:2003ny,Ji:2004gf,Belitsky:2005qn,Feldmann:2007zz,Boffi:2007yc}
and references therein).  Experimentally, the GPDs of the pion are
elusive quantities, as they appear in exclusive processes which are
difficult to measure, such as the deeply virtual Compton scattering or the
hard electro-production of mesons.  Recently, it has been suggested to
study instead the deeply virtual Compton scattering on a virtual pion that
is emitted by a proton~\cite{Amrath:2008vx} under the operating conditions
which will first be met after the energy upgrade at TJLAB. This will eventually 
set important constraints on the pion GPDs.

Despite their fundamental and general character, GPDs are genuinely
defined in the Minkowski space, thus hindering direct determinations on
Euclidean lattices.  The moments of GPDs in the $X$ variable (see
Sect.~\ref{sec:basic}) form polynomials in the $\xi$ variable, with
coefficients depending on the $t$-variable only. The lowest moments
yield the standard electromagnetic and the gravitational form factors,
while higher moments are known as {\em generalized form factors}.
These are useful quantities which may be computed directly on the
lattice. Presently, apart from the values at $t=0$, the generalized
form factors are not known even from lattice simulations due to
insufficient statistics.  In this work we make predictions for the
first few generalized form factors. Since these objects do not
correspond to conserved currents, they evolve with the QCD scale as
they carry anomalous dimensions. We undertake this perturbative
analysis at leading order and show how the generalized form factors
evolve with the scale. We use the techniques described in detail in
Ref.~\cite{Broniowski:2007si}.

Chiral quark models have proved to properly describe the essential
features of the pionic GPDs and related quantities.  The special case
of the parton distribution function (PDF) has been analyzed in the NJL
model in
Refs.~\cite{Davidson:1994uv,RuizArriola:2001rr,Davidson:2001cc}. The
diagonal GPD in the impact parameter space was obtained in
\cite{Broniowski:2003rp}.  Other calculations of the pionic GPDs and
PDFs were performed in instanton-inspired chiral quark models
\cite{Dorokhov:1998up,Polyakov:1999gs,Dorokhov:2000gu,Anikin:2000th,Anikin:2000sb,Praszalowicz:2002ct,Praszalowicz:2003pr,Bzdak:2003qe}.
In the effective quark models the GPDs have been analyzed
in~\cite{Polyakov:1999gs,Theussl:2002xp,Bissey:2003yr,Noguera:2005cc,Broniowski:2007si}.
Studies paying particular attention to polynomiality were carried out
in \cite{Polyakov:1999gs,Tiburzi:2002kr,Tiburzi:2002tq}, which proceeded
via double distributions~\cite{Radyushkin:1998bz}. The same technique
was applied in Ref.~\cite{Broniowski:2007si}, which allowed for a
simple proof of polynomiality in chiral quark models. 
Also, other formal features, such as
support, normalization, crossing, and the positivity
constraints~\cite{Pire:1998nw,Pobylitsa:2001nt}, are all satisfied in
the chiral quark model calculation of Ref.~\cite{Broniowski:2007si}.
Moreover, the obtained analytic expressions have a rather non-trivial
form which is not factorizable in the $t$-variable.  The parton
distribution amplitude (PDA), related to the GPD via a low-energy
theorem \cite{Polyakov:1998ze}, was evaluated in
Refs.~\cite{Anikin:1999cx,Praszalowicz:2001wy,Dorokhov:2002iu,RuizArriola:2002bp,RuizArriola:2002wr}
(see Ref.~\cite{Bakulev:2007jv} for a brief review of analyses of
PDA). The authors of Ref.~\cite{Amrath:2008vx} provide standard
electromagnetic and gravitational form factors based on a
phenomenological model factorization assumption which is generally not
satisfied by our dynamical field theoretical
calculation~\cite{Broniowski:2007si}.  Finally, the related quantity,
the pion-photon transition distribution amplitude (TDA)
\cite{Pire:2004ie,Pire:2005ax,Lansberg:2006fv,Lansberg:2007bu} has been obtained in
Refs.~\cite{Tiburzi:2005nj,Broniowski:2007fs,Courtoy:2007vy,Courtoy:2008af,Kotko:2008gy}.

In the present work we find that in the considered class of models a
simple relation between the gravitational and charge form factors
holds (see the Appendix). The essential element for the proof is the
existence of the spectral representation of the quark model, which is
the case both for SQM and NJL with the Pauli-Villars
regularization. The relation implies that in the considered models and
in the chiral limit the gravitational mean squared radius is half the
electromagnetic one.

In Ref.~\cite{Broniowski:2007si} we have stressed the relevance of the
QCD evolution for the phenomenological success of the chiral quark
models in the description of the experimental and lattice data for the
PDF and PDA of the pion. Indeed, the evolved results for the valence
PDF compare very well to the Drell-Yan data from the E615 experiment
\cite{Conway:1989fs} at the scale of 4~GeV, and to the transverse
lattice results \cite{Dalley:2002nj} at lower scales,
$\sim$0.5~GeV. Similarly, for the case of the PDA the QCD evolution
leads to a fair description of the E791 dijet data \cite{Aitala:2000hb}
at the scale of 2~GeV, and of the transverse lattice
data~\cite{Dalley:2002nj} at the lattice scale.  The comparison is
presented in Figs.~8-11 of Ref.~\cite{Broniowski:2007si}.

The above-mentioned success of the chiral quark models in describing
properties related to the pionic GPD allows us to hope that also their
other aspects, such as the form factors, including the generalized
ones, can be reliably estimated in these models.  Certainly, the
electromagnetic form factor, as one of the most basic quantities, has
been promptly computed in all chiral quark models of the pion. Note
that these evaluations assume the large-$N_c$ limit, thus the effects
of the pion loops, important at low momenta, are absent. We note that
the NJL model leads to somewhat too small electromagnetic radius
\cite{RuizArriola:2002wr}, while in SQM, which incorporates the vector
meson dominance, one may fit the data accurately. Some predictions of
the NJL model for the generalized form factors of the pion have been
presented by one of us in Ref.~\cite{Broniowski:2008dy}, however, the
data away from the physical pion mass have been used in that work.

The outline of the paper is as follows: In Sec.~\ref{sec:basic} we
review the necessary formalism providing basic definitions and
notation. The essential information on the two chiral quark models
used in our work is provided in Sec.~\ref{sec:qm}, while
Sec.~\ref{sec:ff} shows their predictions for the electromagnetic and
gravitational form factors. In Sec.~\ref{sec:lat}, containing our main
results, we compare our predictions to the full-QCD lattice data from
Refs.~\cite{Brommel:2005ee,Brommel:PhD}.  Then in Sec.~\ref{sec:hoff}
we pass to the results for the higher-order form factors, where the
QCD evolution effects are incorporated.
  
\section{Basic definitions \label{sec:basic}}

In this Section we review the basic concepts necessary for our
analysis and introduce the notation.

Throughout this paper we work for simplicity in the strict chiral
limit,
\begin{eqnarray}
m_\pi=0. \label{chiral}
\end{eqnarray}
Since the lattice data of Refs.~\cite{Brommel:2005ee,Brommel:PhD}
perform the extrapolation to the physical pion mass which is small,
the assumption (\ref{chiral}) is appropriate. Nevertheless, the
extension to the physical pion mass in chiral quark models is
straightforward.  In any case, it is worth mentioning that although an
attempt was made~\cite{Brommel:2005ee,Brommel:PhD} to incorporate
chiral logarithms from Chiral Perturbation Theory ($\chi$PT) to the
one-loop order~\cite{Gasser:1983yg,Gasser:1984gg,Gasser:1984ux} (for a
review see e.g. Ref.~\cite{Pich:1995bw}), as described in
Ref.~\cite{Diehl:2005rn}, the data did not exhibit their presence when
all other uncertainties were considered, suggesting instead a linear
extrapolation.

In this paper we will deal with standard gravitational and vector form
factors. The corresponding quark operators are
\begin{eqnarray}
\Theta^{\mu \nu}  = \sum_{q=u,d, \dots }\bar q(x) \frac{\rm i}2 \left( \gamma^\mu \partial^\nu +
\gamma^\nu \partial^\mu \right) q(x) \, , 
\end{eqnarray} 
and 
\begin{eqnarray}
J^{\mu}_V  = \sum_{q=u,d, \dots } \bar q(x) \frac{\tau_a}{2}\gamma^\mu  q(x) \, , 
\end{eqnarray} 
respectively.  The gravitational quark form factors of the pion
\cite{Donoghue:1991qv}, $\Theta_1$ and $\Theta_2$, are defined through
the matrix element of the energy-momentum tensor in the one-pion
state,
\begin{eqnarray}
&& \!\!\!\!\!\! \langle \pi^b(p') \mid \Theta^{\mu \nu}(0) \mid \pi^a(p) \rangle = \\ && 
\!\!\!\!\!\!\! \frac{1}{2}{\delta^{ab}}\left [ (g^{\mu \nu}q^2- q^\mu q^\nu)
\Theta_1(q^2)+ 4 P^\mu P^\nu \Theta_2(q^2) \right ], \nonumber
\end{eqnarray}
where $P=\frac{1}{2}(p'+p)$, $q=p'-p$, and $a, b$ are the isospin
indices.  The gravitational form factors satisfy the low-energy
theorem $\Theta_1(0) - \Theta_2(0) = {\cal O} (m_\pi^2)
$~\cite{Donoghue:1991qv}. In the low momentum and pion-mass limit one
can establish contact with $\chi$PT in the presence of
gravity~\cite{Donoghue:1991qv},
\begin{eqnarray} 
\Theta_1 (q^2 ) \!\!&=&\!\! 1 + \frac{2q^2}{f^2} ( 4 L_{11} + L_{12} ) -
\frac{16 m^2}{f^2} ( L_{11} - L_{13} ) + \dots, \nonumber \\
\Theta_2 (q^2 ) \!\!&=&\!\! 1 - \frac{2 q^2}{f^2} L_{12} + \dots,
\label{eq:theta-low}
\end{eqnarray}
where $f = 86~{\rm MeV}$ is the pion weak decay constant in the chiral
limit and  $L_{11,12,13}$ are the corresponding gravitational low
energy constants.  The calculation of the effective energy-momentum
tensor as well as the low energy constants for chiral quark models has
been carried out in Refs.~\cite{Megias:2004uj,Megias:2005fj}. 
The vector form factor is defined as
\begin{eqnarray}
\langle \pi^a (p')| J_V^{\mu, b} (0) | \pi^c (p)\rangle = \epsilon^{abc}
\left(p'^\mu + p^\mu \right) F_V (q^2) \, , 
\end{eqnarray}
At low momentum one has 
\begin{eqnarray}
F_V (q^2) = 1 + \frac{2 q^2 L_9}{f^2}+ \dots 
\end{eqnarray}
where $L_9$ is a low energy constant of  $\chi$PT~\cite{Gasser:1984ux}.

The generalized quark form factors of the pion (here we take $\pi^+$
for definiteness) are defined as matrix elements of quark bilinears
accompanied by additional derivative operators,
\begin{eqnarray}
 \label{eq:gff}
 && \bra{\pi^{+}(p')} \overline{u}(0)\, \gamma^{\{\mu}\, i \lrd\/^{\mu_1}
  i \lrd\/^{\mu_2} \dots i \lrd\/^{\mu_{n-1}\}} \,u(0) \ket{\pi^{+}(p)} = \nonumber \\
 && \;\;\;\;2 P^{\{\mu}P^{\mu_1} \dots P^{\mu_{n-1}\}} A_{n0}(t)
    +  \\ && \;\;\;\;2\sum^n_{\substack{k=2\\\textrm{even}}}q^{\{\mu}
    q^{\mu_1} \dots q^{\mu_{k-1}}
    P^{\mu_{k}} \dots P^{\mu_{n-1}\}} \,2^{-k}A_{nk}(t), \nonumber
\end{eqnarray}
where $A_{nk}(t)$ are the generalized form factors with $n=1,2,\dots$
and $k=0,2,\dots,n$. The symbol $\rd$ is the QCD covariant derivative,
$\lrd = \frac{1}{2} (\rd -\ld)$, and $\{\dots\}$ denotes the
symmetrization of indices and the subtraction of traces for each
pair of indices. The factor of $2^{-k}$ is conventional and makes our
definition different than in
Refs.~\cite{Brommel:2005ee,Brommel:PhD}. For $n=1$ and $2$ we have
\begin{eqnarray}
&& A_{10}(t)=F_V(t), \nonumber \\
&& A_{20}(t)=\frac{1}{2}\Theta_1^q(t), \;\; A_{22}(t)=-\frac{1}{2}\Theta_2^q(t), \label{lowff}
\end{eqnarray}
where $F_V$ is the pion electromagnetic form factor, while
$\Theta_i^q$ denote the quark parts of the gravitational form factors
of Eq.~(\ref{eq:gff}). 

Chiral quark models work at the scale where the only explicit degrees of  
freedom are quarks, while the gluons are integrated out. Thus, 
the gluon form factors of the pion vanish, 
\begin{eqnarray}
A^G_{nk}(t)=0 \;\; ({\rm quark-model~scale}). \label{lowffG}
\end{eqnarray}
When evolution to higher scales is carried out
\cite{Broniowski:2007si}, non-zero gluonic moments are generated.
Note that the above mentioned soft-pion theorem
$\Theta_1(0)=\Theta_2(0) + {\cal O}(m_\pi^2)$ applies to the {\it
  full} trace of the energy momentum tensor, and consequently does not
apply to their quark or gluonic contributions separately. The
condition (\ref{lowffG}) translates then into $A_{20}(0)=-A_{22}(0)$ in the
chiral limit at the quark model scale.

Now we pass to the definition of the GPDs, whose moments are related
to the form factors listed above.  The used conventions are the same
as in Ref.~\cite{Broniowski:2007si}, where also all the details of the
calculation of the GPDs can be found. The kinematics of the GPDs can
be read off from Fig.~\ref{fig:diag}. The adopted notation is
\begin{eqnarray}
&& p^2=m_\pi^2=0, \;\; q^2=-Q^2=-2p\cdot q=t,\nonumber \\ && n^2=0, \;\; p
\cdot n=1, \;\; q \cdot n=-\zeta, \label{kin}
\end{eqnarray}
with the null vector $n$ defining the light cone.  The two isospin
projections of the quark GPDs of the pion, isosinglet (singlet) and
isovector (non-singlet), are defined through the matrix elements of
quark bilinears, with the quark fields displaced along the light cone,
\begin{eqnarray}
&& \delta_{ab}\,{\cal H}^{I=0}(x,\zeta,t) = \int \frac{dz^-}{4\pi}
e^{i x p^+ z^-}  \\ && \;\;\; \times \left . \langle \pi^b (p+q) | \bar
\psi (0) \gamma \cdot n \psi (z) | \pi^a (p) \rangle \right
|_{z^+=0,z^\perp=0}, \nonumber \\ && i \epsilon_{3ab}\,{\cal
H}^{I=1}(x,\zeta,t) = \int \frac{dz^-}{4\pi} e^{i x p^+ z^-} 
\label{defGPD01} \\ && \;\;\; \times \left . \langle \pi^b (p+q) | \bar \psi
(0) \gamma \cdot n \psi (z) \, \tau_3 | \pi^a (p) \rangle \right
|_{z^+=0,z^\perp=0}. \nonumber
\end{eqnarray} 
The coordinate $z$ lies on the light cone.  At the quark-model scale
the gluon GPD of the pion vanishes,
\begin{eqnarray}
{\cal H}^{G}(x,\zeta,t)=0 \;\; ({\rm quark-model~scale}).
\end{eqnarray}
When the QCD evolution to higher scales is performed, a non-vanishing
gluonic contribution ${\cal H}^{G}(x,\zeta,t)$ is
generated~\cite{Broniowski:2007si}.

In the following analysis we use the {\em symmetric} notation for the GPDs,
\begin{eqnarray}
\xi= \frac{\zeta}{2 - \zeta}, \;\;\;\; X = \frac{x - \zeta/2}{1 -\zeta/2}, \label{xiX}
\end{eqnarray}
where $0 \le \xi \le 1$ and the support is $-1 \le X \le 1$. One introduces the corresponding GPDs
\begin{eqnarray}
H^{I=0,1}(X,\xi,t)={\cal H}^{I=0,1}\left ( \frac{\xi + X}{\xi + 1},
\frac{2 \xi}{\xi + 1},t \right ).
\end{eqnarray}
The reflection about $X=0$ yields the symmetry relations
\begin{eqnarray}
H^{I=0}(X,\xi,t)&=&-H^{I=0}(-X,\xi,t), \nonumber \\
H^{I=1}(X,\xi,t)&=&H^{I=1}(-X,\xi,t). \label{eq:sym}
\end{eqnarray} 
At $X \ge 0$ the GPDs are related to the PDF, $q(X)$, 
\begin{eqnarray}
{H}^{I=0}(X,0,0)= {H}^{I=1}(X,0,0)= q(X). \nonumber
\label{eq:q} 
\end{eqnarray}

The {\em polynomiality} conditions~\cite{Ji:1998pc,Radyushkin:2000uy}
follow from very basic field-theoretic assumptions, namely the Lorentz
invariance, time reversal, and hermiticity, which yields the 
form~(\ref{eq:gff}). For the moments one finds 
\begin{eqnarray}
\int_{-1}^1 \!\!\!\!\! dX\,X^{2j} \, {H}^{I=1}(X,\xi,t) = 2\sum_{i=0}^j A_{2j+1,2i}(t) \xi^{2i}, \nonumber \\
\int_{-1}^1 \!\!\!\!\! dX\,X^{2j+1} \, {H}^{I=0}(X,\xi,t) = 2\sum_{i=0}^{j+1} A_{2j+2,2i}(t) \xi^{2i}, \label{poly}
\end{eqnarray}
with $j=0,1,\dots$. For the lowest moments one has
\begin{eqnarray}
&&\int_{-1}^1 \!\!\!\!\! dX\,        {H}^{I=1}(X,\xi,t) = 2A_{10}(t)=2F_V(t), \label{norm} \\
&&\int_{-1}^1 \!\!\!\!\! dX\, X   \, {H}^{I=0}(X,\xi,t) = 2A_{20}(t)+2A_{22}(t)\xi^2 \nonumber \\
&& \hspace{3.5cm}= \Theta_2(t)-\Theta_1(t) \xi^2. \label{norm2} 
\end{eqnarray}
In the convention of Refs.~\cite{Brommel:2005ee,Brommel:PhD} the
equivalent expansion is in powers of $(2\xi)^2$ rather than $\xi^2$,
as in Eq.~(\ref{poly}).  In our approach, used in the following
Sections polynomiality is explicitly manifest from the use of the
double distributions \cite{Broniowski:2007si}.  The equivalence of
Eq.~(\ref{eq:gff}) and (\ref{poly}) is easily proven by contracting
(\ref{eq:gff}) with the null vectors $n^{\mu_1} \dots n^{\mu_j}$ and
subsequently applying the definitions (\ref{kin}).  We notice that for
the isovector GPD only the even, and for the isoscalar GPD only the
odd moments are non-vanishing.  The gluon form factors are defined as the
integrals
\begin{eqnarray}
\int_{-1}^{1}dX\,X^{2j+1} H_g(X,\xi,t,Q^2)=2\sum_{i=0}^{j+1} A^G_{2j+2,2i}(t)\,\xi^{2i}.\nonumber \\
\end{eqnarray}

Finally, we note that Eq.~(\ref{poly}) for $j=0$ expresses the
electric charge conservation and the momentum sum rule operating in
deep inelastic scattering.

\section{Chiral quark models \label{sec:qm}}

\begin{figure}[tb]
\subfigure{\includegraphics[width=5.8cm]{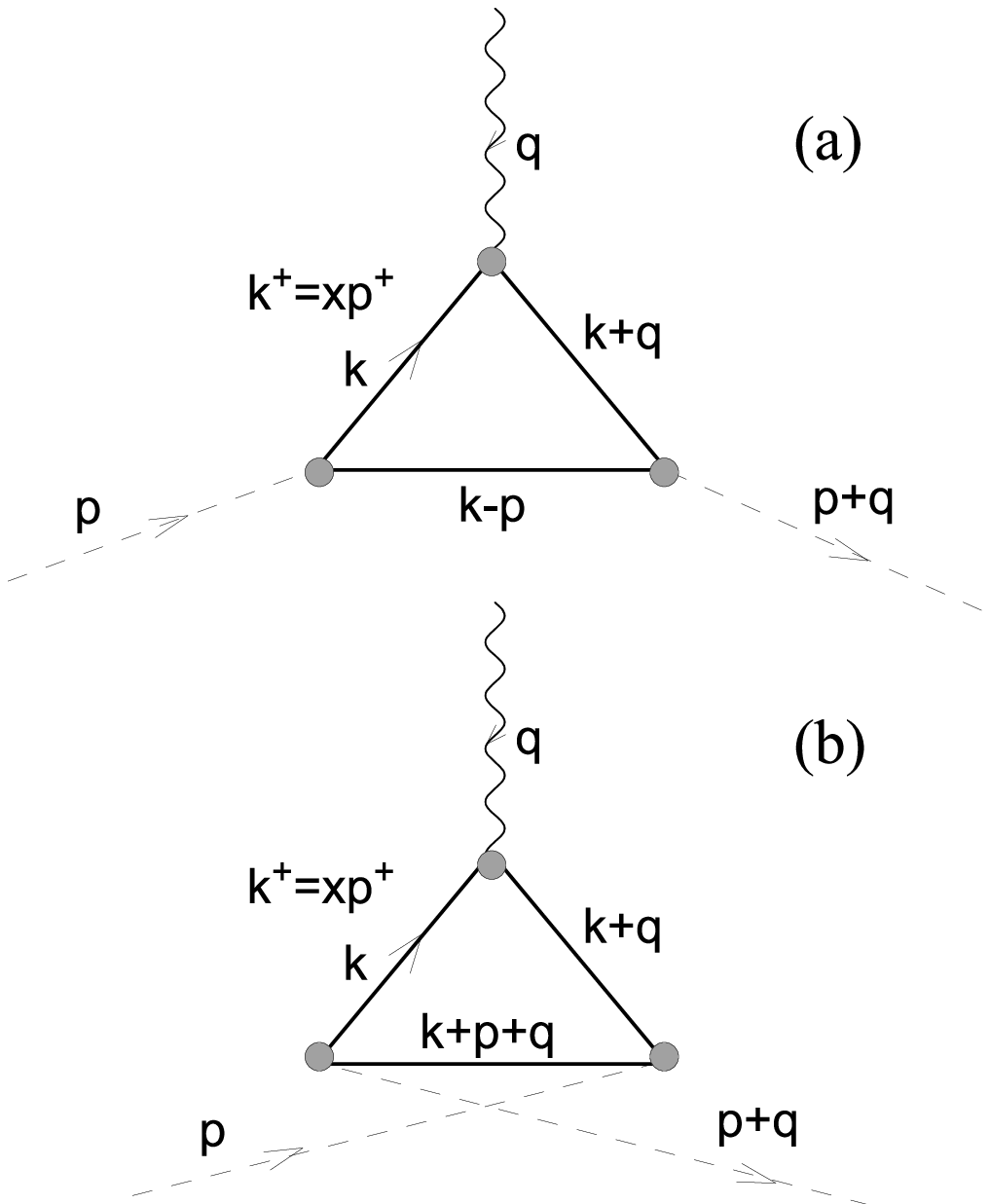}}\\
\subfigure{\includegraphics[width=3.7cm]{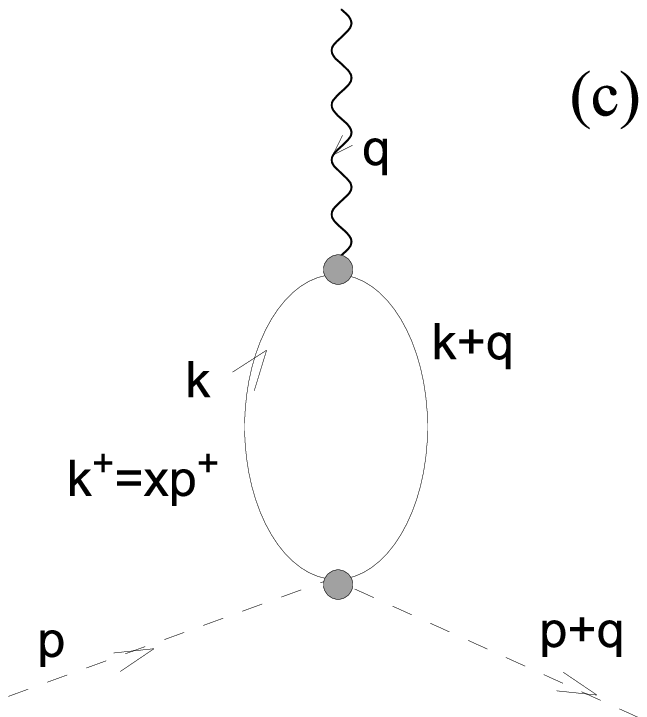}} 
\vspace{-2mm}
\caption{(Color online) The direct (a), crossed (b), and contact (c) Feynman diagrams for the
quark-model evaluation of the GPD of the pion. The contact contribution is responsible for the $D$-term.}
\label{fig:diag} 
\end{figure}
In chiral quark models at the leading-$N_c$ level the calculation of
the form factors and GPDs proceeds according to the one-loop diagrams
of Fig.~\ref{fig:diag}. Extensive details of the quark-model
evaluation are given in \cite{Broniowski:2007si}.  In this paper we
carry out calculations in two chiral quark models: SQM
\cite{RuizArriola:2003bs} and NJL with the Pauli-Villars
regularization in the twice-subtracted version of
Refs.~\cite{RuizArriola:1991gc,Schuren:1991sc, RuizArriola:2002wr}. Variants of chiral quark models differ
in the way of performing the necessary regularization of the quark
loop diagrams.

The spectral quark model \cite{RuizArriola:2003bs} introduces the
generalized spectral density $\rho(\omega)$ in the quark mass
$\omega$, in the spirit of Ref.~\cite{Efimov:1993zg}, supplied with
chiral symmetry, gauge invariance, and vector meson dominance. The
one-quark-loop action of SQM has the form
\begin{eqnarray}
\Gamma_{\rm SQM} =-i N_c \int_C d \omega \rho(\omega) {\rm Tr} \log
\left ( i\slashchar{\partial} - \omega U^5 \right),
\label{eq:eff_ac} 
\end{eqnarray} 
where $\rho(\omega)$ is the quark generalized {\em spectral function},
and $U^5=\exp( i \gamma_5 \vec{\tau} \cdot \vec{\phi} /f)$,  with
$\vec{\phi}$ denoting the pion field in the nonlinear realization.
The vector part of the spectral function, needed in the present
analysis, has the {vector meson-dominance} form
\cite{RuizArriola:2003bs}
\begin{eqnarray}
\rho_V (\omega) &=& \frac{1}{2\pi i} \frac{1}{\omega}
\frac{1}{(1-4\omega^2/m_\rho^2)^{5/2}},  \label{rhov}
\end{eqnarray}
exhibiting the pole at the origin and cuts starting at $\pm m_\rho/2$,
where $m_\rho\sim 770$~MeV is the mass of the rho meson.  The complex
contour $C$ for the integration in (\ref{eq:eff_ac}) is given in
Ref.~\cite{RuizArriola:2003bs}.  SQM leads to conventional and
successful phenomenology for both the pion
\cite{RuizArriola:2003bs,RuizArriola:2003wi,Megias:2004uj,Broniowski:2007si},
the nucleon \cite{Arriola:2006ds}, and the photon parton distribution
amplitude \cite{Dorokhov:2006qm}.

We also study a more conventional chiral quark model, the NJL model
with the Pauli-Villars regularization in the twice-subtracted version
proposed in
Refs.~\cite{RuizArriola:1991gc,Schuren:1991sc,RuizArriola:2002wr}. The
one-quark-loop action of the model is
\begin{eqnarray}
\Gamma_{\rm NJL} =-i N_c {\rm Tr} \log
\left( i\slashchar{\partial} - M U^5 \right),
\label{eq:eff_ac_NJL} 
\end{eqnarray} 
where $M$ is the constituent quark mass. The Pauli-Villars
regularization is introduced at the effective action 
level \cite{RuizArriola:1991gc,Schuren:1991sc, RuizArriola:2002wr}, with the
practical advantage that gauge and relativistic symmetries as well as
sum rules are manifestly fulfilled~\cite{Weigel:1999pc}. For the
observables considered in this paper the Pauli-Villars is implemented
according to the prescription, where instances of $M^2$ in an
observable ${\cal O}$ are replaced with $M^2+\Lambda^2$, and then the
regularized observable is evaluated according to the prescription
\begin{eqnarray}
{\cal O}_{\rm reg} = {\cal O}(0) - {\cal O}(\Lambda^2 ) + \Lambda^2
\frac{d {\cal O}(\Lambda^2 )}{d\Lambda^2}. \label{prescr}
\label{eq:PV2} 
\end{eqnarray} 
The Pauli-Villars regulator $\Lambda$ is a free parameter of the
model.  In what follows we use
\begin{eqnarray}
M=280~{\rm MeV}, \;\;\; \Lambda=871~{\rm MeV}, \label{njlpar} 
\end{eqnarray}
which yields $f=93.3$~MeV for the pion decay constant
\cite{RuizArriola:2002wr} according the the formula
\begin{eqnarray}
f^2=-\frac{N_c M^2}{4\pi^2} \left [ \log(\Lambda^2 + M^2) \right ]_{\rm
  reg}. \label{f2njl}
\end{eqnarray}

\section{Electromagnetic and gravitational form factors \label{sec:ff}}

The form factors may be calculated in two different ways. The first
method uses the definition (\ref{eq:gff}), which leads to the
evaluation of one-loop diagrams with an appropriate vertex. For the
electromagnetic form factor the vertex is $Q\gamma^\mu$, where $Q$ is
the electric charge of the quark.  For the gravitational form factor
the vertex, corresponding to the energy-momentum tensor, has the form
\begin{eqnarray}
&&\Theta^{\mu \nu}(k+q,k)=\frac{1}{4} \left [ (2k+q)^\mu \gamma^\nu +
    (2k+q)^\nu \gamma^\mu \right ] \nonumber \\ && - \frac{1}{2}
  g^{\mu \nu} \left ( 2 \slashchar{k} + \slashchar{q} - \omega \right
  ),
\end{eqnarray}
with $\omega$ denoting the quark mass. We illustrate the calculation
in the Appendix. The other method uses the GPDs obtained
earlier~\cite{Broniowski:2007si} and evaluates their moments
(\ref{poly}). The results are the same, which serves as a consistency
test of the algebra.  As mentioned in Sec.~\ref{sec:basic}, the equivalence is proven
by contracting with the null vector.  For instance in the case of the
gravitational form factor we consider $n_\mu \Theta^{\mu \nu}
n_\nu$. Then
\begin{eqnarray}
&& \langle \pi^b(p+q) \mid n_\mu \Theta^{\mu \nu}(0) n_\nu \mid \pi^a(p) \rangle = \\ && 
\delta^{ab} \frac{1}{2}\left [
\zeta^2 \Theta_1(q^2)+ (2-\zeta)^2 \Theta_2(q^2) \right ] \nonumber
\end{eqnarray}
and the vertex becomes 
\begin{eqnarray}
&&n_\mu \Theta^{\mu \nu}(k+q,k)n_\nu =(x-\zeta/2) \gamma \cdot n.  
\end{eqnarray}
We recognize the same vertex as in the evaluation of the GPD's
multiplied by $(x-\zeta/2)$.  Upon passing to the symmetric notation
Eq.~(\ref{norm2}) follows.

Before showing the explicit results both for SQM and NJL models in the
specific realizations described in the previous section, we note some
general results. Actually, in the considered quark models and in the
chiral limit we have the following identity relating the gravitational
and electromagnetic form factor,
\begin{eqnarray}
\frac{d}{dt} \left[ t\, \Theta_i (t ) \right] &=& F_V(t ) \, , \quad (i=1,2) \, , 
\label{eq:FV-theta}
\end{eqnarray}
from which the identity between the two gravitational form factors
$\Theta_1 (t ) = \Theta_2 (t ) \equiv \Theta(t)$ follows.  This
remarkable quark model relations are proven explicitly in the
Appendix. The essential ingredient of the proof is the existence of
the spectral representation in both considered models.  One
consequence of relation (\ref{eq:FV-theta}) is the expected
consistency of normalizations at $t=0$ for the charge and mass
$\Theta_i(0 ) = F_V(0)$, displaying the tight connection between the
gauge and Poincare invariances. Furthermore, expanding in small $t$
and using the fact that $F(t)=F(0) [ 1- \langle r^2 \rangle t/6 +
  \dots] $ with $\langle r^2 \rangle $ denoting the mean squared
radius, one gets
\begin{eqnarray}
2 \langle r^2 \rangle_\Theta = \langle r^2 \rangle_V, 
\label{eq:r2-V-theta}
\end{eqnarray} 
which means that in the considered models and in the chiral limit
the gravitational mean squared radius is half the electromagnetic one.

Turning now to the specific realization, by construction, the pion
electromagnetic form factor in SQM has the monopole form
\begin{eqnarray}
F_V^{\rm SQM}(t)=\frac{m_\rho^2}{m_\rho^2-t}. \label{ffSQM}
\end{eqnarray}
A straightforward evaluation for the  gravitational form factor yields
\begin{eqnarray}
\Theta_1^{\rm SQM}(t)=\Theta^{\rm SQM}_2(t)=\frac{m_\rho^2}{t} \log
\left ( \frac{m_\rho^2}{m_\rho^2-t} \right ) \equiv \Theta^{\rm
  SQM}(t). \nonumber \\ \label{ffSQMg}
\end{eqnarray}
 These specific form factors fulfill trivially the general relations
 Eq.~(\ref{eq:FV-theta}) and Eq.~(\ref{eq:r2-V-theta}).

In the NJL model the pion electromagnetic form factor is equal to
\begin{eqnarray}
&&F_V^{\rm NJL}(t)=1+ \frac{N_c g_\pi^2}{8 \pi^2}  \label{ffnjl} \left ( 
\frac{2 s \log \left(\frac{s-\sqrt{-t}}{s+\sqrt{-t}}\right)}{\sqrt{-t}}
\right )_{\rm reg}, \nonumber
\end{eqnarray}
with $g_\pi=M/f$ denoting the quark-pion coupling constant. We have introduced the 
short-hand notation $s=\sqrt{4 \left(M^2+\Lambda ^2\right)-t}$.
The condition $\lim_{t \to - \infty} F_{\rm NJL}(t)=0$ is satisfied due to
Eq.~(\ref{f2njl}).  
For the gravitational form factor we obtain 
\begin{eqnarray}
\!\!\!\!\!\!\!&&\Theta^{\rm NJL}(t)=F_V^{\rm NJL}(t)-\frac{N_c g_\pi^2}{8 \pi^2} \times \label{ffnjlg} \\
\!\!\!\!\!\!\!&&\left ( 
\frac{4 (M^2+\Lambda^2)}{t} \left[ {\rm Li}_2 \left ( \frac{2\sqrt{-t}}{s+\sqrt{-t}} \right )+
{\rm Li}_2 \left ( \frac{2\sqrt{-t}}{\sqrt{-t}-s}\right ) \right ]
\right )_{\rm reg}, \nonumber
\end{eqnarray} 
where ${\rm Li}_2(z)$ is the polylogarithm function. 

Of course, the explicit expressions (\ref{ffSQM},\ref{ffSQMg}) and
(\ref{ffnjl},\ref{ffnjlg}) comply to the general relation
(\ref{eq:FV-theta}).

Similar expressions, of growing complication, may be obtained for
higher-order form factors in both SQM and NJL models.

\section{Comparison to the lattice data \label{sec:lat}}

\begin{figure}[tb]
\subfigure{\includegraphics[width=.48\textwidth]{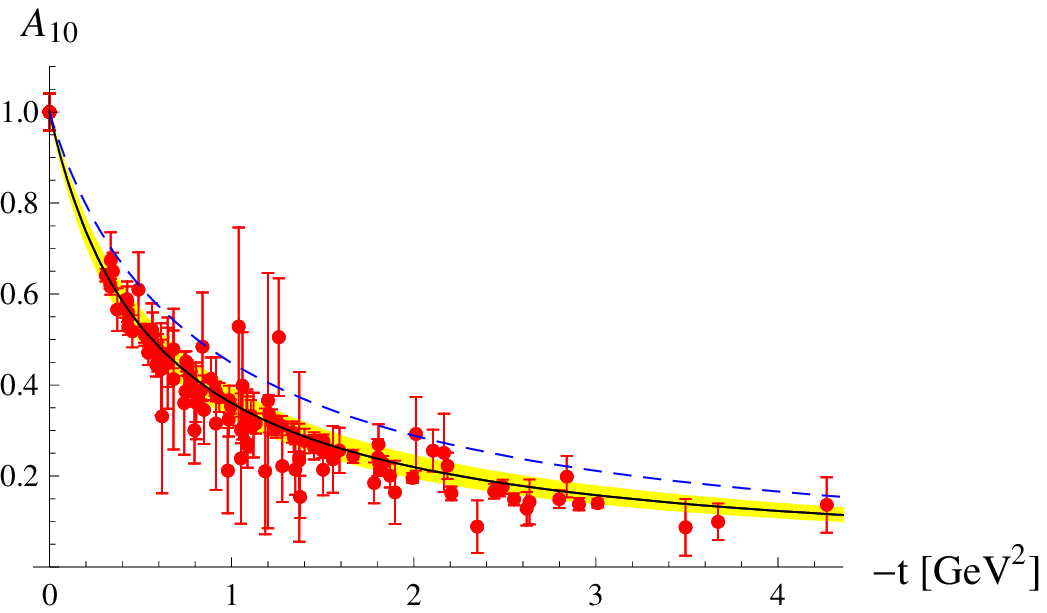}} \\
\subfigure{\includegraphics[width=.48\textwidth]{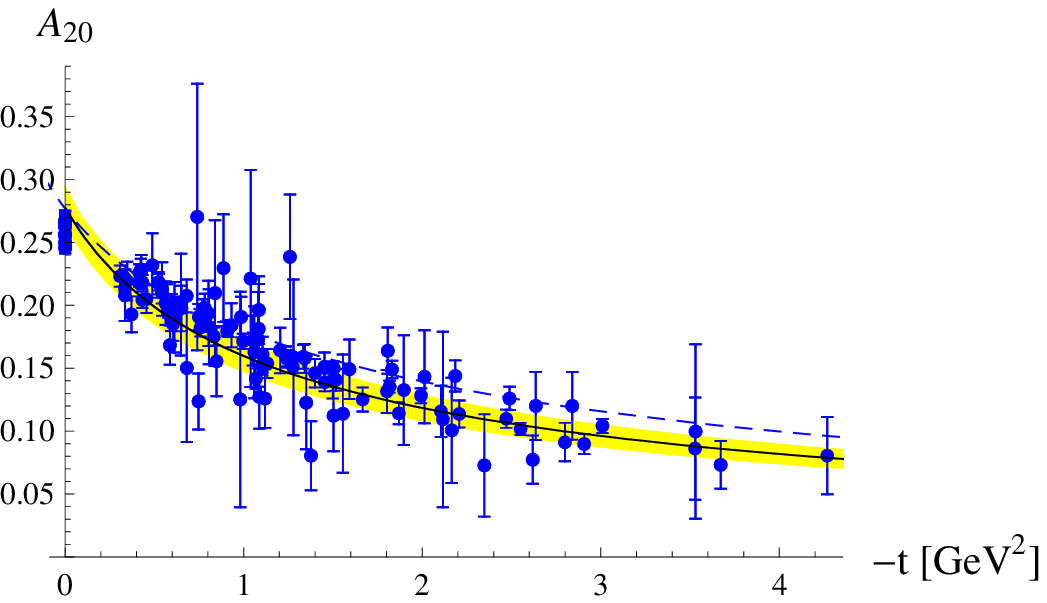}} 
\vspace{-2mm}
\caption{(Color online) The electromagnetic form factor (top) and
  quark part of the gravitational form factor (bottom) in SQM (solid
  line) and NJL model (dashed line) compared to the lattice data from
  Ref.~\cite{Brommel:PhD}. The band around the SQM results corresponds
  to the uncertainty in the quark momentum fraction $R$ and the
  $m_\rho$ parameter. \label{fig:gv}}
\end{figure}

\begin{figure}[tb]
\subfigure{\includegraphics[width=.48\textwidth]{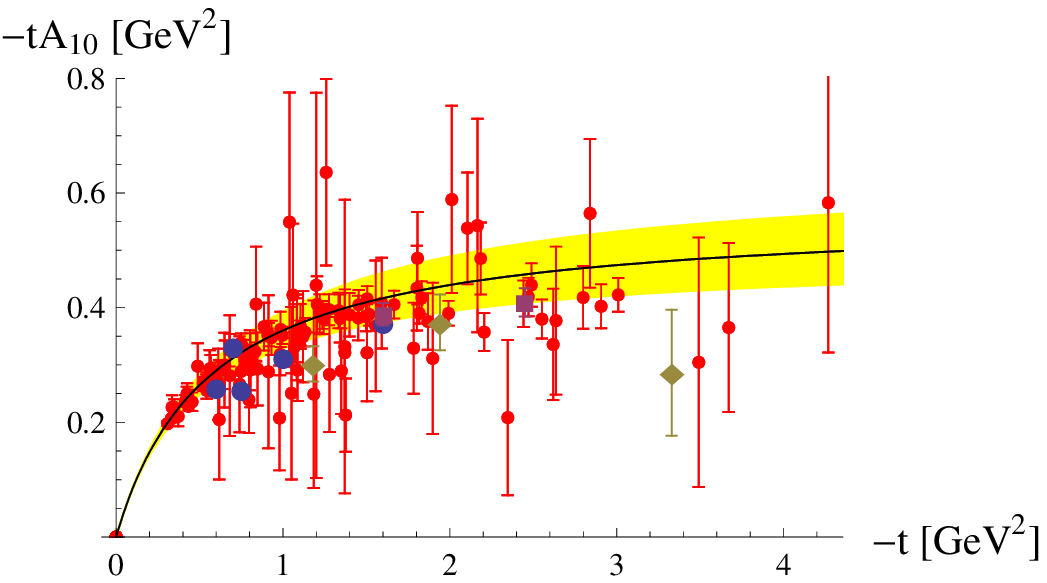}} \\
\subfigure{\includegraphics[width=.48\textwidth]{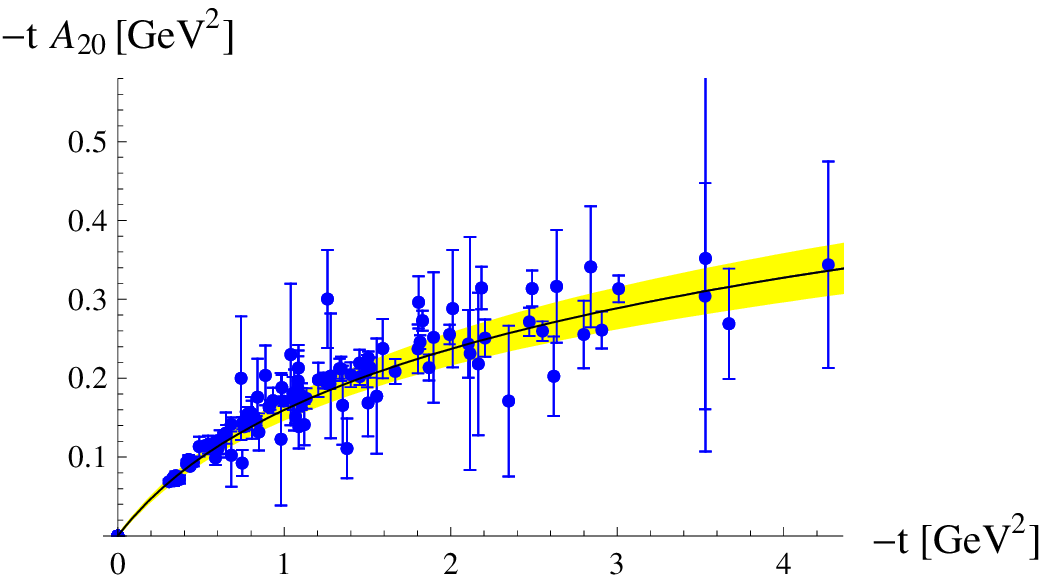}} 
\vspace{-2mm}
\caption{(Color online) Same as Fig.~{\ref{fig:gv} for the form
    factors multiplied with $-t$. In addition, we include the TJLAB
    data \cite{Volmer:2000ek,Tadevosyan:2007yd,Horn:2006tm} (darker
    and larger circles and squares) and the Cornell data
    \cite{Bebek:1977pe} (diamonds).} \label{fig:gvt}}
\end{figure}

We are now ready to compare the results of the previous Section to the
recent lattice data \cite{Brommel:2005ee,Brommel:PhD}.  These data
correspond to the scale of $Q=2$~GeV and give the quark part of the
gravitational form factor and the electromagnetic form factor.  On the
other hand, the quark model calculation corresponds to a very low
scale, $Q_0\sim 320$~MeV and in general the QCD evolution is needed to
compare the model predictions to the data at a different scale $Q$.

A detailed discussion of the evolution issue is presented in Ref.~\cite{Broniowski:2007si}.
The low energy Chiral Quark Models contain {\em quarks} as the basic and explicit
degrees of freedom and the considered twist-2 observables
correspond to modeling QCD at a low renormalization
point. Obviously, by the energy-momentum conservation in these models the quarks as the only degrees of freedom 
carry 100 \% of the total momentum in a hadron. On the other hand the momentum fraction
carried by the valence quarks in the pion at the scale $Q^2= 4 {\rm GeV}^2$ is 
about 40\%.  The QCD evolution (LO, NLO, N2LO) tells us that this number grows as the scale 
is evolved to {\em lower values}. The quark model
reference scale $Q_0$ is determined as the scale where the evolved QCD
value yields exactly the inescapable 100\% of the quark model. It
would of course be highly desirable to include, {\em e.g.}, explicit gluonic
degrees of freedom as this would allow to stop the perturbative evolution
at a higher scale. Nonetheless, the LO and NLO evolutions for the PDF's
are sufficiently close to each other \cite{Davidson:2001cc} as to provide some
confidence on the kind of calculations carried out in our work.

The issue of the QCD evolution of GPDs and the generalized form
factors is addressed in detail in Sect.~\ref{sec:hoff}. However, the
electromagnetic and gravitational form factors do not evolve with the
scale. What changes is the ratio $R$ of the total momentum fraction
carried by the quarks (valence and sea)
\begin{eqnarray}
R=\frac{ \langle x \, \rangle_q (Q) } { \langle x \, \rangle_q (Q_0)  } = \left( \frac{\alpha(Q)}
{\alpha(Q_0) } \right)^{\gamma_1^{\rm (0)} / (2 \beta_0) } \quad ,
\qquad
\end{eqnarray} 
where the anomalous dimension is given by $ \gamma_1^{\rm (0)} / (2
\beta_0) = 32/81 $ for $N_F=N_c=3$ and
\begin{eqnarray}
&&\alpha(Q^2)=\frac{4\pi}{\beta_0 \log(Q^2/\Lambda^2_{QCD})} \\
&&\beta_0=\frac{11}{3}N_c-\frac{2}{3}N_f, \label{gambe}
\end{eqnarray}
where we take $\Lambda_{\rm QCD} = 226~{\rm MeV}$ and $N_c=N_f=3$.  At
the scale $Q_0$ we have $R=1$, which then gradually decreases with the
increasing scale.  The quark part of the gravitational form factor is
$\Theta^q(t)=R \Theta(t)$, the gluon part is $\Theta^G(t)=(1-R)
\Theta(t)$, such that, of course, $\Theta^q(t)+\Theta^G(t)=\Theta(t)$.

The value of $R$ depends on the evolution ratio
$\alpha(Q^2)/\alpha(Q_0^2)$, which is not precisely known on the
lattice. For that reason we shall treat $R$ as a free parameter when
fitting the model results to the data for the gravitational form factor.
In SQM the other parameter is the value of the rho meson mass,
$m_\rho$. For SQM we fit jointly the electromagnetic and the quark
part of the gravitational form factor. The $\chi^2$ method yields
\begin{eqnarray}
R=0.28\pm 0.02, \;\;\; m_\rho=0.75\pm 0.05~{\rm GeV}, \label{par}
\end{eqnarray}
with $\chi^2/{\rm DOF}=1.8$. The result of the fit is displayed in
Fig.~\ref{fig:gv} with the solid line. The band corresponds to the
uncertainties in the values of parameters in Eq.~(\ref{par}). We note
an overall very good agreement for both form factors. Note a
significantly slower fall-off for the gravitational form factor
compared to the electromagnetic one.  The optimum value of $m_\rho$
agrees within the error bars, which are substantial, with the physical
mass of the rho meson.

For the case of the NJL model (dashed line in Fig.~\ref{fig:gv}) the
agreement is not as good as in the SQM model and it is not possible to
improve it by changing the parameters $M$ and $\Lambda$. For that
reason we have not carried out the $\chi^2$ fit in this case.  The
problems of the NJL model in reproducing the electromagnetic form
factor, where the corresponding rms radius turns out to be too small,
are well known, see {\em e.g.} the discussion in
Ref.~\cite{RuizArriola:2002wr}. Similar discrepancy can be noted for
the case of the gravitational form factor shown in the bottom panel of
Fig.~\ref{fig:gv}.

Now we come back in a greater detail to the issue of the quark to
gluon momentum ratio $R$. It evolves with the scale from the value
$R=1$ at the quark model scale to $R \to 0$ at $Q^2 \to \infty$. We use the
standard LO DGLAP evolution for the single channel. We know the final
scale, $Q=2$~GeV, but the quark model scale is a priori unknown. We
thus adjust $Q_0$ in order to reach the value of $R$ from
Eq.~(\ref{par}) at the known scale $Q$.  The result is
\begin{eqnarray}
Q_0=0.31\pm 0.03~{\rm GeV} \label{scale}
\end{eqnarray}
with $\Lambda_{\rm QCD}=226~{\rm MeV}$.
This value agrees with the earlier independent determinations of the
quark-model scale, see Ref.~\cite{Broniowski:2007si} for a detailed
discussion on these issues.  We note that at such low scales as in
Eq.~(\ref{scale}) the perturbative expansion parameter in the DGLAP
equations is large, ${\alpha(Q^2_0)}/({2\pi})=0.34$, which makes the
evolution very fast for the scales in the vicinity of $Q_0$.

One should note that the large value of $\alpha(Q^2_0)$ calls for the
use of improved QCD evolution at low scales.  The presented analysis
could be extended in several ways, for instance incorporating the
next-to-leading (NLO) effects, or by modifying the dependence of
$\alpha$ on $Q^2$ at low scales incorporating the ``infra-red
protection'' \cite{Shirkov:1997wi}.  The NLO corrections to the meson
PDF's in chiral quark models were studied in
Ref.~\cite{Davidson:2001cc}, where, somewhat surprisingly, it was
found that these effects are small. Modifications in the evolution of
$\alpha(Q^2)$ could be incorporated along the lines of
Refs.~\cite{Stefanis:2000vd,Bakulev:2004cu,Bakulev:2005fp} where they
were used for the pion form factor, but this calculation is outside of
the scope of the present paper.

We stress that the chiral quark model explanation of the lattice data
used in this work as well as the physical and lattice data explored in
Ref.~\cite{Broniowski:2007si} requires ``strong'' evolution, with a
large value of the evolution ratio $\alpha(Q_0^2)/\alpha(Q^2)$.  Our
simple-minded analysis based on the LO evolution may probably be
viewed as an approximation to a more elaborate scheme, nevertheless it
is quite remarkable that various observables (PDF, PDA, their moments)
lead to the same evolution ratio, giving in the LO approximation
compatible quark-model scales \cite{Broniowski:2007si}. Moreover, the
electromagnetic and the gravitational form factors are independent of
the evolution, hence the issue does not arise for these observables.

In Fig.~\ref{fig:gvt} we show the form factor multiplied by $-t$,
which is a popular way to present the results at large Euclidean
momenta. In the case of the vector (electromagnetic) form factor we
also display the experimental TJLAB data
\cite{Volmer:2000ek,Tadevosyan:2007yd,Horn:2006tm} and the earlier
Cornell data \cite{Bebek:1977pe} and the approximately constant value
for $-t F_V(t)$ is clearly seen. In the gravitational case, however,
the lattice data~\cite{Brommel:2005ee,Brommel:PhD} show an increasing
trend which is well mimicked by our SQM form factor, $- t \Theta(t)
\sim \log (-t)$. 

In the large-$N_c$ limit in the single resonance approximation (SRA)
\cite{Donoghue:1991qv,Pich:2002xy} one would have a monopole form
factor
\begin{eqnarray}
\Theta^{\rm SRA} (t ) =  \frac{A \, M_f^2}{M_f^2-t} \, , 
\end{eqnarray}
which agrees with our quark model at low $t$ if $M_f = \sqrt{2} M_V
\sim 1100 {\rm MeV}$ in agreement with the result of
Eq.~(\ref{eq:r2-V-theta}). The difference between the SRA form factor
and our Eq.~(\ref{ffSQMg}) is less than $10\%$ for momentum values up
to $ t= -1 {\rm GeV}^2 $. The SRA monopole fits well the data
extrapolated to the  physical pion mass, yielding $A=0.261(5)$ and $M_f
= 1320(60){\rm MeV}$~\cite{Brommel:2005ee,Brommel:PhD}.

To conclude this Section, several general remarks are in order.  One
has to bear in mind that the full-QCD lattice calculations are
linearly extrapolated to the physical pion mass. Our quark-model
analysis incorporates only the leading-$N_c$ contributions, while the
full-QCD simulations include all orders in that expansion.
Nevertheless, despite these caveats, we note a quite remarkable
agreement, in particular for the case of SQM. This is noteworthy, as
our study is a genuine dynamical field-theoretical calculation, and
the parameters providing the optimal fit are highly compatible with
calculations of other processes within the {\em same model and scheme}.

Admittedly, at very large values of $-t$ perturbative QCD results for
the form factors should be reached. 
At very low values of $t$ chiral corrections are
important. Since the data we use are at intermediate values of
Euclidean momenta, we may neglect both above-mentioned effects and use
the chiral quark models to explain the data.
 
We stress that our calculation conforms to the low-energy theorem
$\Theta_1(0) - \Theta_2(0) = {\cal O}
(m_\pi^2)$~\cite{Donoghue:1991qv} which is dictated by chiral
symmetry. The quark model predicts, in addition,
$\Theta_1(t)=\Theta_2(t)$, and the purely multiplicative character of
the QCD evolution yields in our conventions
\begin{eqnarray}
A_{22}(t)=-A_{20}(t).
\end{eqnarray}
Probably due to insufficient statistics, this formula is not quite
seen in the data of Ref.~\cite{Brommel:PhD} (note that with
conventions adopted in that reference one should have instead 
$A_{22}(t)=-\frac{1}{4} A_{20}(t)$). For that reason we have not used
the data for $A_{22}$ in our numerical analysis.

\section{Higher-order form factors \label{sec:hoff}}

\begin{figure}[tb]
\subfigure{\includegraphics[width=.48 \textwidth]{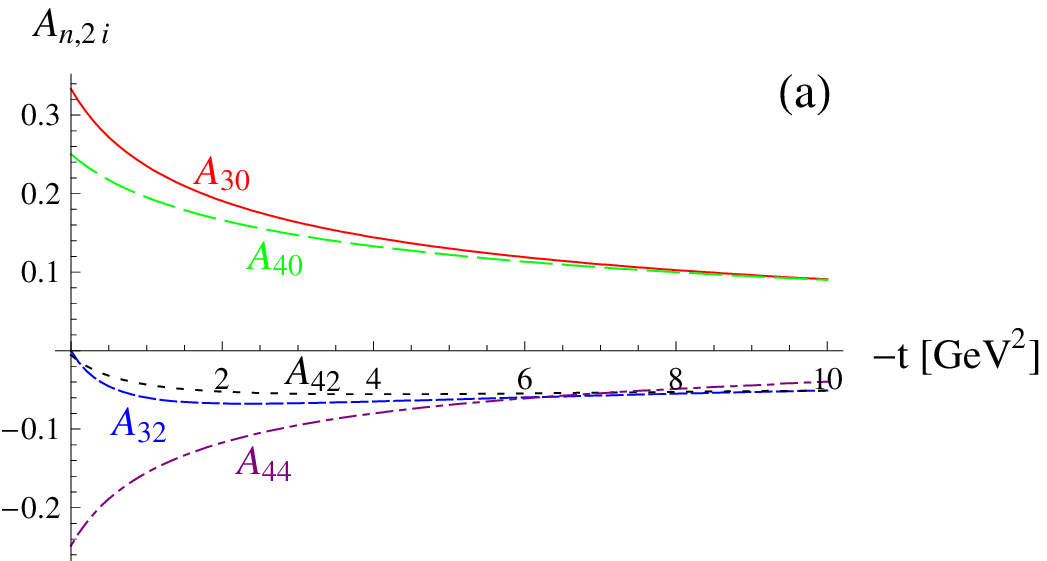}}\\
\subfigure{\includegraphics[width=.48 \textwidth]{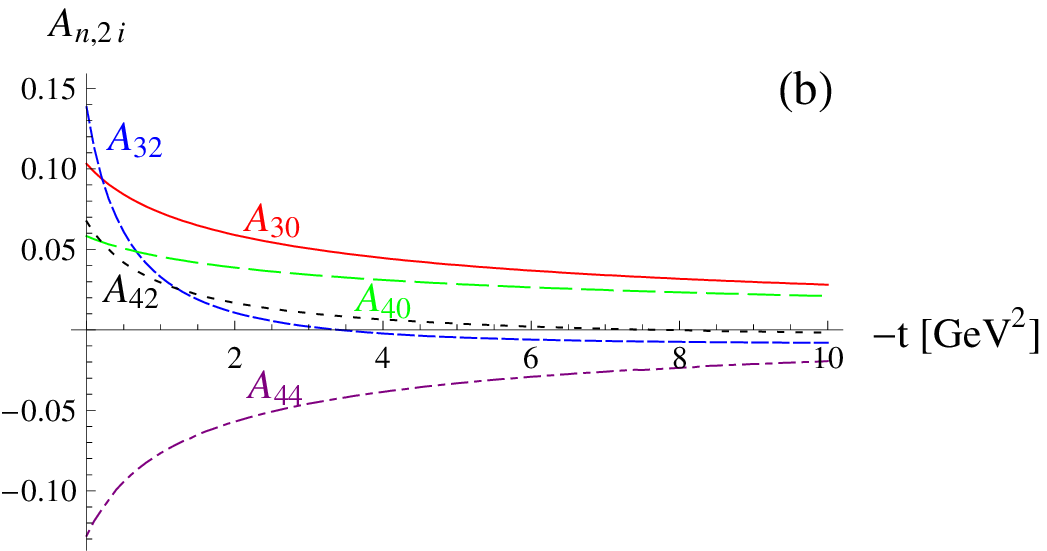}}\\
\subfigure{\includegraphics[width=.48 \textwidth]{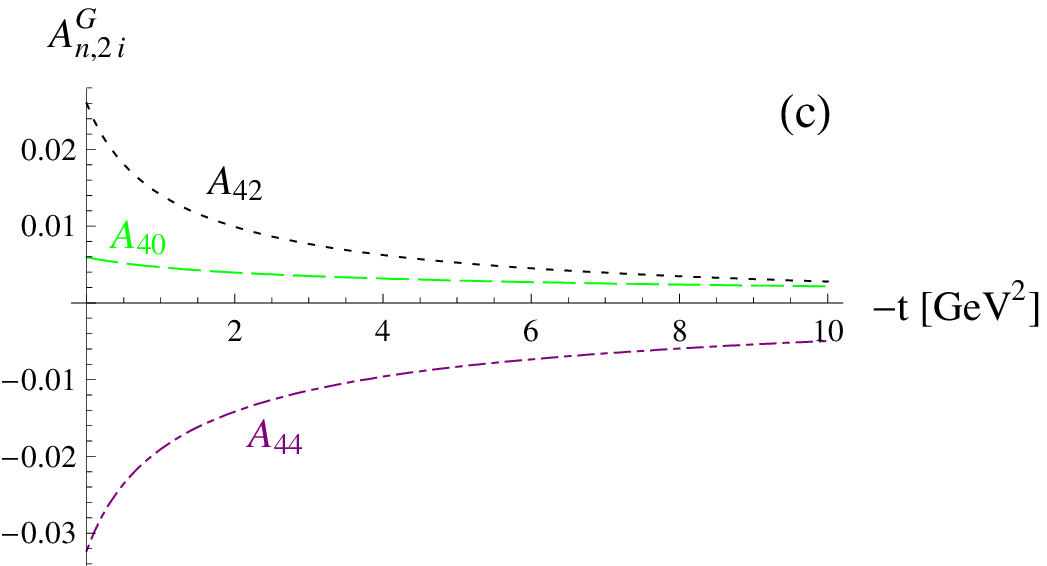}}
\vspace{-2mm}
\caption{(Color online) Generalized form factors $A_{3,2i}$ and
  $A_{4,2i}$ of the pion in SQM at
  the quark-model scale $Q_0$ (a), at the lattice scale $Q=2$~GeV (b),
  and the gluon form factors $A^G_{4,2i}$ at
  $Q=2$~GeV(c). \label{fig:mom34}}
\end{figure}

Analogous calculations as in the previous section can be performed for
the higher-order generalized form factors.  Here we only give the
results for the case of SQM, as the results in NJL are qualitatively
similar, while also SQM works better for those quantities which can be
confronted to the data.  In Fig.~\ref{fig:mom34}(a) we show $A_{3,2i}$
and $A_{4,2i}$ obtained at the quark model scale.  In the chiral
quark models in the chiral limit one has at $t=0$ very simple
expressions \cite{Polyakov:1999gs,Theussl:2002xp,Broniowski:2007si}
\begin{eqnarray}
&& {H}^{I=1}(X,\xi,0) = \theta \left(1-X^2\right)  \\
&& {H}^{I=0}(X,\xi,0) =  \nonumber \\
&& \;\;\;\; \theta ((1-X) (X-\xi))-\theta ((X+1) (-\xi-X)),\nonumber
\end{eqnarray}
where $\theta(x)$ is the Heavyside step function. It follows from the
definition (\ref{poly}) that at the quark-model scale the following
relations hold \cite{Broniowski:2008dy},
\begin{eqnarray}
A_{2j+1,2i}(0)&=& \left \{ \begin{array}{cl} \frac{1}{2j+1} & {\rm for}\; i=0 \\ 0 & {\rm otherwise}\end{array}\right . \nonumber \\
A_{2j+2,2i}(0)&=& \left \{ \begin{array}{cl} \frac{1}{2j+2} & {\rm for}\; i=0 \\ -\frac{1}{2j+2} & {\rm for}\; i=j+1 \\ 
0 & {\rm otherwise}\end{array}\right . \nonumber \\
&& \hspace{1.3cm}{\rm (quark-model~scale)}
\end{eqnarray}
These relations can be seen  in Fig.~\ref{fig:mom34}(a). 
The form factors tend to zero very slowly at large $-t$. 

Another property follows from the fact that in the considered model
${H}^{I=0}(X,1,t)=0$ for any value of $t$. Then Eq.~(\ref{poly})
yields \cite{Broniowski:2008dy}
\begin{eqnarray}
\sum_{i=0}^{j+1} A_{2j+2,2i}(t)=0.
\end{eqnarray}
This feature can be seen in the lower panel of Fig.~\ref{fig:mom34}.

While the lowest order form factors, {\em i.e.} the electromagnetic
and the gravitational, do not evolve with the running scale $Q$, the
higher-order form factors change when we pass from the reference quark
model scale $Q_0$ to $Q$. Our analysis proceeds as follows: We carry out
the leading-order DGLAP-ERBL evolution for the quark-model GPDs from
the quark-model scale
\begin{eqnarray}
Q_0=0.31~{\rm GeV}
\end{eqnarray} 
to the scale of the lattice calculation, $Q=2$~GeV. Note that the GPDs
at the quark-model scale do not exhibit factorization in the
$t$-variable.  In fact, this is the reason for the non-trivial change
of the higher-order form factors with the scale.  After the evolution
we take the moments (\ref{poly}) at several values of $\xi$ and then
disentangle the generalized form factors via solving a set of linear
equations.  For the LO ERBL DGLAP evolution we use the method and the
numerical program of Ref.~\cite{GolecBiernat:1998ja}.  The result of
our calculations is presented in Fig.~\ref{fig:mom34}(b).  We notice a
dramatic change, both in the value at $t=0$ and in shape of the form
factors compared to the behavior of Fig.~\ref{fig:mom34}(a) at the
quark-model scale.

Formally, as $Q^2\to \infty$ the GPDs approach their asymptotic
forms contained entirely in the ERBL region $|X|<\xi$. Explicitly, we
have in this limit\cite{Broniowski:2007si}
\begin{eqnarray}
\label{eq:asy}
&&H^{I=1}= \frac{3}{2\xi} \left(1-\frac{X^2}{\xi^2}\right) F_V(t) 
\\
&&H^{I=0}= (1 - \xi^2) \frac{15}{4 \xi^2} \frac{N_f}{4 C_F + N_f} \frac{X}{\xi}\left(1-\frac{X^2}{\xi^2}\right)\Theta(t) \nonumber
\\
&&X H^G= (1 - \xi^2)  \frac{15}{4 \xi} \frac{C_F}{4C_F + N_f} \left(1-\frac{X^2}{\xi^2}\right)^2 \Theta(t),\nonumber\\
&& \hspace{21mm} (Q^2\to \infty) \nonumber 
\end{eqnarray} 
where $C_F=(N_c^2-1)/(2N_c)$ and $N_f=3$.
The proportionality factors follow from the normalization
at the initial quark-model scale $Q_0$, as the  charge- and momentum-conservation sum rules
are invariants of the evolution,
\begin{eqnarray}
&&\int_{-1}^{1}dX\,H^{I=1}(X,\xi,t,Q^2)=2F_V(t),\\
&&\int_{-1}^{1}dX\,\left[ X H^{I=0}(X,\xi,t,Q^2)+X H_g(X,\xi,t,Q^2)\right] \nonumber \\
&&=(1-\xi^2) \Theta(t), \nonumber
\end{eqnarray}
in accordance to Eq.~(\ref{norm},\ref{norm2}).

Evaluation of moments in Eq.~(\ref{eq:asy}) yields immediately
\begin{eqnarray}
A_{2j+1,2i}(t)&=& \left \{ \begin{array}{cl} \frac{3}{4j(j+2)+3} F_V(t) & {\rm for}\; i=j 
                                                      \\ 0 & {\rm otherwise}\end{array}\right . \nonumber \\
A_{2j+2,2i}(t)&=& \left \{ \begin{array}{cl} \frac{N_f}{4 C_F + N_f}\frac{15}{2[4j(j+4)+15]} \Theta(t) & {\rm for}\; i=j 
                                                      \\ -A_{2j+2,2j}(t) & {\rm for}\; i=j+1 \\
                                                                      0 & {\rm otherwise}\end{array}\right . \nonumber \\
A^G_{2j+2,2i}(t)&=& \frac{4C_f}{N_f} \frac{1}{2j+1} A_{2j+2,2i}(t), \nonumber \\
&& \hspace{21mm} (Q^2\to \infty) \label{asGFF}                                                              
\end{eqnarray}
We note a qualitative difference of the asymptotic form factors
compared to the form factors at the quark model scale shown in
Fig.~\ref{fig:mom34}.  For the isovector case ($n=2j+1$) only the
highest form factor, with $i=j$, does not vanish, while for the
isoscalar case ($n=2j+2$) only the two highest moments, with $i=j+1$
and $i=j$ are non-zero. The remaining moments tend to zero. This
result is a prompt conclusion from the asymptotic forms
(\ref{eq:asy}).  Note that in contrast to this behavior, at the
quark-model scale all generalized form factors are non-zero.  As
mentioned previously, the form factors $A_{10}$, $A_{20}$, and
$A_{22}$ are invariants of the evolution.  The asymptotic gluon form
factors in Eq.~(\ref{asGFF}) are related to the isoscalar quark form
factors in a simple manner.  Asymptotically, all generalized form
factors become proportional to $F_V(t)$ or $\Theta(t)$ in the
isovector and isoscalar channels, respectively.

Finally, we compare our values of the higher-order form factors at
$t=0$ to the lattice data provided in Sec.~7 of
Ref.~\cite{Brommel:PhD}.  With the notation for the moments at $t=0$,
\begin{eqnarray}
\langle x^n \rangle = A_{n+1,0}(0),
\end{eqnarray}
one finds at the lattice scale of $Q=2$~GeV
\begin{eqnarray}
&&\langle x \rangle = 0.271\pm 0.016, \\
&& \langle x^2 \rangle = 0.128\pm 0.018, \nonumber\\
&& \langle x^3 \rangle = 0.074\pm 0.027. \nonumber \\
&& \hspace{3cm} (\rm lattice) \nonumber
\end{eqnarray} 
while in both chiral quark models we obtain after the LO DGLAP
evolution to the lattice scale
\begin{eqnarray}
&&\langle x \rangle = 0.28\pm 0.02, \\
&& \langle x^2 \rangle = 0.10\pm 0.02, \nonumber \\
&& \langle x^3 \rangle = 0.06\pm 0.01, \nonumber \\
&& \hspace{.8cm} (\rm chiral~quark~models) \nonumber
\end{eqnarray}
where the error bars come from the uncertainty of the scale $Q_0$ in
Eq.~(\ref{scale}).  The two sets of numbers overlap within the error
bars. This result is quite remarkable, as it shows that the hierarchy
of the form factors at $t=0$ found in full-QCD lattice calculations is
properly reproduced in chiral quark models.  The above form factors
are simply the moments of the PDF of the pion. We recall that the PDF
itself in the chiral quark models reproduces very well the
experimental \cite{Conway:1989fs} and transverse lattice data of
Ref.~\cite{Dalley:2002nj}, see Figs.~8 and 9 of
Ref.~\cite{Broniowski:2007si}.

\section{Conclusions \label{sec:concl}}

Here are our main points:

\begin{enumerate}

\item The gravitational form factor obtained from the spectral quark
  model in the chiral limit agrees very well with the lattice data of
  Refs.~\cite{Brommel:2005ee,Brommel:PhD}. We have performed a global
  fit to the electromagnetic and the quark part of the gravitational
  form factors and obtained very reasonable values for the evolution
  ratio and the vector-meson mass.  The longer range of the
  gravitational form factor follows from a different analytic
  expression compared to the electromagnetic form factor.
 
\item The NJL model does not provide such an excellent agreement,
  although the qualitative features are very similar to SQM.

\item We provide analytic expressions for the lowest-order form
  factors in both considered models. For the considered models in the
  chiral limit we find an explicit relation between the gravitational
  and vector form factors. In particular, the relation shows that
  in our case both gravitational form factors are equal, and that the
  mean squared electromagnetic radius is twice the gravitational
  one. 

\item The electromagnetic and gravitational form factors do not evolve
  with the scale, but the higher-order generalized form factors do.
  We have performed the leading-order ERBL DGLAP QCD evolution of the
  pion GPDs and obtained via moments the generalized form factors at
  the scale of the lattice measurements.

\item The generalized form factors at $t=0$ found in full-QCD lattice
  simulations are reproduced in chiral quark models within the error
  bars corresponding to statistical and model uncertainties.

\item Our predictions can be further tested with future lattice
  simulations for higher-order form factors. The behavior is
  non-trivial, with form factors having different signs, magnitude,
  and asymptotic fall-off.

\item Lattice simulations for the gluon form factors would provide a
  very useful independent information, which could be used to verify
  the model predictions.

\end{enumerate}

\begin{acknowledgments}
The authors are grateful to Dirk Br\"ommel for helpful e-mail exchanges
and for providing the lattice data of Refs.~\cite{Brommel:PhD} for our
figures, as well as cordially thank Krzysztof Golec-Biernat for making available
his computer code for solving the QCD evolution equations for the
GPDs.
\end{acknowledgments}

\appendix

\section{Relation between gravitational and electromagnetic form factor}

In this Appendix we show an explicit calculation of form factors in
SQM and prove as a by-product the result of Eq.~(\ref{eq:FV-theta}).
The proof also applies to the NJL model with the PV regularization, as
that model can also be written in terms of the spectral representation
(see below).

The SQM is defined by the generalized Lehman representation of the quark
propagator
\begin{eqnarray}
S(\p) = \int_C d \omega \frac{ \rho( \omega )}{ \slashchar{p} -
\omega} \, .  
\label{eq:lehmann} 
\end{eqnarray}
The method exploited in Ref.~\cite{RuizArriola:2003bs} was mainly
based in writing down Ward identities for the corresponding vertex
functions at the quark level, and then closing the quark line to make
one loop calculations. As discussed in later works (see
e.g. \cite{Megias:2004uj,Arriola:2006ds}) this is fully equivalent to
proceeding directly through the effective action which we sketch now.
Using the action of Eq.~(\ref{eq:eff_ac}) one can compute the energy
momentum tensor as a functional derivative of the action with respect
to an external space-time-dependent metric, $ g_{\mu\nu} (x)$, around
the flat space-time metric $ \eta_{\mu \nu} $ (we take the signature
$(+ - - -)$ )
\begin{eqnarray}
&&\frac12 \Theta^{\mu \nu} (x) = \frac{\delta \Gamma}{\delta g_{\mu
    \nu}(x)} \Big|_{g_{\mu \nu} = \eta_{\mu\nu} } \\ && \;\; = -{\rm i}
\frac{N_c}2 \int_C d \omega \rho(\omega) \langle x | \left\{ O^{\mu
  \nu} \, , \, \left(i\slashchar{\partial} - \omega U^5 \right )^{-1}
\right\} | x \rangle \, , \nonumber
\label{eq:theta_quark}
\end{eqnarray} 
where
\begin{eqnarray}
O^{\mu \nu} = \frac{\rm i}2 \left( \gamma^\mu \partial^\nu +
\gamma^\nu \partial^\mu \right) -  g^{\mu\nu} \left({\rm i}
\slashchar{\partial}- \omega \right) \, ,  
\end{eqnarray} 
and $U^5=\exp( i \gamma_5 \vec{\tau} \cdot \vec{\phi} /f)$ with
$\vec{\phi}$ the pion field in the nonlinear realization.  For the
calculation of the gravitational form factor, we expand up to second
order in the pion field corresponding to evaluating the diagrams of
Fig.~\ref{fig:diag}. In the cartesian isospin basis one has
\begin{eqnarray}
&& \!\!\!\! \langle \pi^b |\Theta_{\mu\nu} | \pi^a \rangle = -N_c 
\int d\omega 
\rho(\omega ) \int \frac{d^4 k}{(2\pi)^4 } {\rm
Tr} \Big\{ \Theta_{\mu \nu} (k+q,k )  \nonumber \\ && \times \Big[ {i\over \slashchar{k} - \omega } \left( -\frac{\tau_a
\gamma_5 \w}f \right) {i\over \slashchar{k}-\slashchar{p} - \omega }
\left( -\frac{\tau_b \gamma_5 \w }f \right) {i\over
\slashchar{k}+\slashchar{q} - \omega } \nonumber \\ && \phantom{pepe}
+ {i\over \slashchar{k} - \omega } \frac{i \delta_{ab} \w}{ 2
f^2}{i\over \slashchar{k}+\slashchar{q} - \omega } \Big] + {\rm
crossed} \Big\} \, , 
\end{eqnarray} 
where the quark gravitational vertex is given by
\begin{eqnarray}
 \Theta_{\mu \nu} (k+q,k ) &=& \frac14 \Big[ (2k_\mu + q_\mu)
 \gamma_\nu + (2k_\nu + q_\nu) \gamma_\mu \Big] \nonumber \\ &-&
 \frac12 g_{\mu \nu} ( 2{\slashchar{k}}+{\slashchar{q}} - \omega ) \Big] \, .
\end{eqnarray} 
After computing the traces and using the Feynman trick in
the integrals, in the chiral limit the result is
\begin{eqnarray}
\Theta_1 (q^2 ) &=& \frac{N_c}{4 \pi^2 f^2} \int d \omega
\rho(\omega) \omega^2 \int_0^1 \frac{dx}{x(1-x) q^2 } \nonumber
\\&\times& \Big[ - \omega^2 \log \omega^2 + x(1-x) q^2 \nonumber \\ &+&(
\omega^2 - x(1-x) q^2) \log ( \omega^2 - x(1-x) q^2) \Big] \, ,\nonumber \\ 
\Theta_2 (q^2 ) &=& \Theta_1 (q^2 ) \, .
\label{eq:theta12}
\end{eqnarray} 
In the low-momentum limit we may use Eq.~(\ref{eq:theta-low}) as
deduced from  $\chi$PT 
in the presence of gravity~\cite{Donoghue:1991qv} to get 
\begin{eqnarray} 
L_{11} &=& \frac{N_c}{192 \pi^2} \, , \\ L_{12} &=&
-\frac{N_c}{96 \pi^2} \, , \\ 
L_{13} &=& -\frac{N_c}{(4\pi)^2} \frac{{\rho_1}' }{12 B_0} =
\frac{f^2}{24 M_S^2}  \, , 
\end{eqnarray} 
in agreement with the derivative expansion of
SQM~\cite{Megias:2004uj}.

Proceeding similarly with the vector form factor we get 
\begin{eqnarray}
&& 
\langle \pi^a | J_V^{\mu, b} | \pi^c \rangle  
= -N_c \int d\omega \rho(\omega ) \int
  {d^4 k \over (2\pi)^4 } {\rm Tr} \Big[
    \gamma_\mu \frac{\tau_b}2
\nonumber \\ && \times {i\over \slashchar{k} - \omega } \left(
    -\frac{\tau_c \gamma_5 \w }f \right) {i\over
      \slashchar{p}+\slashchar{k} - \omega } \left( -\frac{\tau_a
      \gamma_5 \w }f \right) {i\over \slashchar{q}-\slashchar{k} -
      \omega } \Big] \, . \nonumber \\ 
\end{eqnarray} 
For on-shell massless pions the electromagnetic form factor reads
\begin{eqnarray}
F_V (q^2 ) &=& -\frac{N_c}{4\pi^2 f^2} \int d\omega \rho(\omega)
\omega^2 \nonumber \\ &\times& \int_0^1 dx \log\left[\omega^2 - x(1-x) q^2 \right] \, .
\label{eq:ffpi}
\end{eqnarray}
Charge normalization, $F_V(0)=1$, and energy-momentum normalization
$\Theta_2(0)=1$ implies 
\begin{eqnarray}
1 = F_V (0 ) = \Theta_2(0) &=& -\frac{N_c}{4\pi^2 f^2} \int d\omega \rho(\omega)
\omega^2 \log \omega^2 \nonumber \\ &=& \frac{N_c m_\rho^2 }{24 \pi^2 f^2 }\, .
\end{eqnarray}
where in the second line the vector meson realization,
Eq.~(\ref{rhov}), has been used. The value agrees with the value of
the pion weak decay constant obtained from the corresponding axial
matrix element~\cite{RuizArriola:2003bs}.

With the representations of Eq.~(\ref{eq:theta12}) and
Eq.~(\ref{eq:ffpi}) the result in Eq.~(\ref{eq:FV-theta}) can be
readily derived, without any
reference to the specific realization given by  
Eq.~(\ref{rhov}). 

The above proof also holds for the NJL model in the PV regularization.
This class of models can be cast explicitly in the
spectral-representation form, using $\rho(\omega)=\delta(\omega-M) +
\sum_i c_i \delta(\omega -\sqrt{M^2+\Lambda_i^2})$, where $c_i$ and
$\Lambda_i$ are the PV constants.  Then all above algebraic steps
carry over and the result (\ref{eq:FV-theta}) holds for the NJL model
with PV regularization as well.

\bibliography{gff}

\end{document}